\documentclass[aps,prd,12pt,nofootinbib,notitlepage]{revtex4-1}

\usepackage{amsfonts}
\usepackage{amsmath,epsfig,bm}
\usepackage{graphicx}
\usepackage{bm}
\usepackage{color}
\renewcommand{\vec}[1]{\mbox{\boldmath$#1$}}

\usepackage{tabularx} 

\usepackage{times}

\newcommand{\be}{\begin{equation}}
\newcommand{\ee}{\end{equation}}
\newcommand{\ba}{\begin{eqnarray}}
\newcommand{\ea}{\end{eqnarray}}
\newcommand{\bd}{\begin{displaymath}}
\newcommand{\ed}{\end{displaymath}}

\def\thalf{{\textstyle{\frac{1}{2}}}}

\def\oneqt{{\textstyle{\frac{1}{4}}}}

\def\rt3{\sqrt{3}}
\def\rt6{\sqrt{6}}
\def\mL2{(m_{\phi}L)^2}

\begin{document}

\title{{\bf Spin-Vorticity Coupling for Massive Vector Mesons}}
\author{Joseph I. Kapusta$^1$, Ermal Rrapaj$^{1,2}$, and Serge Rudaz$^1$}
\affiliation{$^1$School of Physics and Astronomy, University of Minnesota, Minneapolis, Minnesota 55455, USA \\
$^2$Department of Physics, University of California, Berkeley, CA 94720, USA}

\vspace{.3cm}

\parindent=20pt

\begin{abstract}
Recent experiments at Relativistic Heavy Ion Collider (RHIC) and Large Hadron Collider (LHC) have indicated that hadrons containing strange quarks produced in non-central heavy ion collisions can be polarized.  We investigate in detail the coupling of spin and vorticity for electrically neutral, massive vector bosons using the Proca equation, and provide the nonrelativistic reduction of the field equations via a single Foldy--Wouthuysen transformation.  We find that the resulting Hamiltonian is not-Hermitian, but ${\cal PT}$ invariant, and involves a spin dependent term $\thalf s_z \hbar \omega$  to leading order in vorticity.
We also calculate further relativistic and quantum corrections to the Hamiltonian.
\end{abstract}

\date{\today}

\maketitle

\section{Introduction}

Several insightful theoretical papers suggested that the large orbital angular momentum of the matter created in non-central high energy heavy ion collisions could polarize the quarks and subsequently the hadrons observed in the final state \cite{LiangWang1,Betz,Becattini1,Becattini2}.  Measurements by the STAR Collaboration at the Relativistic Heavy Ion Collider (RHIC) of the polarization of the $\Lambda$ and $\bar{\Lambda}$ hyperons were consistent with this idea \cite{FirstSTAR,Nature,SecondSTAR}.  The inferred vorticity $\omega = (9 \pm 1) \times 10^{21}$ s$^{-1}$ is the highest ever measured.  It translates to an energy of $\omega = 6$ MeV (we use units in which $\hbar = c = k_B = 1$).  The observed hyperon polarization decreases with increasing beam energy, becoming nearly zero at the maximum RHIC energy of $\sqrt{s_{NN}} = 200$ GeV.  Measurements of the hyperon polarization by the ALICE Collaboration at the much higher beam energies available at the Large Hadron Collider (LHC) are consistent with zero \cite{LambdaALICE}.
According to the quark model the spin of the $\Lambda$ and $\bar{\Lambda}$ hyperons is carried by the $s$ and $\bar{s}$ quarks.  The important question of how long it takes for the strange quarks to reach and maintain equilibrium with the vorticity was addressed in several papers by the present authors \cite{KRR1,KRR2,KRR3}.

Massive spin-1 vector mesons should also be polarized in non-central high energy heavy ion collisions \cite{LiangWang2,Becattini1,Becattini2,YangWang1,Tang}.  Early, relatively low statistics measurements by the STAR Collaboration at RHIC found no spin alignment of the $K^{*0}(892)$ and $\phi (1020)$ vector mesons \cite{STARvectors}.  Very surprisingly, spin alignment of these vector mesons was measured at the much higher LHC energies by the ALICE Collaboration \cite{KphiALICE}.  (Here it should be noted that, unlike hyperons, the statistical spin density matrix must be used to infer the spin alignments of the vector mesons \cite{DensityMatrix}.)  One possible explanation of this puzzle has been proposed \cite{QunWang2020}.  In addition, ALICE has found no discernable polarization of the $J/\psi$ meson \cite{JpsiALICE}.

Motivated by these experimental results we investigate in detail the coupling of spin and vorticity for electrically neutral, massive vector bosons using the Proca equation.   The outline of our paper is as follows.  Components of the field strength tensor for massive vector fields may be referred to as electric and magnetic fields, even though it is not electromagnetism.  How these fields are defined, whether it be via contravariant or covariant tensors, is reviewed in Sect. \ref{secII}.  The field equations of motion in a rotating frame of reference are presented in a concrete fashion in Sect. \ref{secIII}.  The Hamiltonian for a Schr\"odinger description of the dynamics is presented in Sect. \ref{secIV}.  The nonrelativistic reduction of the field equations via a single Foldy--Wouthuysen transformation is given in Sect. \ref{secV}.  It turns out that the energy states are split as $0, \pm \thalf \omega$, not $0, \pm \omega$ as one might have expected.  It turns out that a non-Hermitian, but ${\cal PT}$ invariant term arises in the nonrelativistic reduction; it is a correction of order $\hbar/c^2$ and so does not appear in classical physics.  It has been shown that ${\cal PT}$ invariant Hamiltonians are not necessarily unphysical.  We explore this term specifically in Sect. \ref{secVI}.  Conclusions are presented in Sect. \ref{secVII}.  Our results may also be relevant to rapidly rotating, cold, trapped atomic gases.  Some details and elaborations are presented in the Appendices.

The description of massive vector mesons in inertial frames of reference, including their interaction with electromagnetic fields if they are charged, is textbook material \cite{WGreiner}.  

\section{Definition of Electric and Magnetic Fields}
\label{secII}

Consider a massive spin-1 vector meson.  The field strength tensors in the inertial and rotating frames of reference are related by $G^{\mu\nu} = e^{\mu}_{\;\; a} e^{\nu}_{\;\; b} \bar{G}^{ab}$ and that the fields are related by $\phi^{\mu} = e^{\mu}_{\;\;a} \bar{\phi}^a$ \cite{KRR3}.  A bar refers to that quantity in the inertial frame, and the $e^{\mu}_{\;\;a}$ are the tetrads \cite{KRR1}.  A simple calculation shows that
\be
\phi^{\mu} \phi_{\mu} = g_{\mu\nu} \phi^{\mu} \phi^{\nu} = \eta_{ab} \bar{\phi}^a \bar{\phi}^b = \bar{\phi}^a \bar{\phi}_a
\ee
Hence the functional form of the Lagrangian ${\cal L} = -\oneqt \bar{G}_{ab} \bar{G}^{ab} + \thalf m^2 \bar{\phi}^a \bar{\phi}_a
= -\oneqt G_{\mu\nu} G^{\mu\nu} + \thalf m^2 \phi^{\mu} \phi_{\mu}$ is unchanged.  See Appendix A for explicit expressions for the metric, the tetrads, and the affine connection.

Concerning the (pseudo) electric and magnetic fields, they can be defined via the contravariant field strength tensor $G^{\mu\nu}$ as $G^{10} = E_x$ and $G^{12} =  - B_z$, etc., or via the covariant field strength tensor $G_{\mu\nu}$ as $G_{01} = E_x$ and $G_{12} =  - B_z$, etc. where ${\bf E} = (E_x, E_y, E_z)$ and 
${\bf B} = (B_x, B_y, B_z)$.  When the metric is $\eta_{\mu\nu}$ it makes no difference  which way they are defined.  Otherwise there is no unique definition of the electric and magnetic fields in the non inertial frame of reference.  Long discussions can be found in Refs. \cite{Schiff,Crater,Ridgely,Osmanov} among many others.  In this paper, the electric and magnetic fields are defined by the contravariant tensor, which results in the covariant components
\ba
G_{0i} &=& \left[ (1-v^2) {\bf E} + ({\bf v} \cdot {\bf E}) {\bf v} + {\bf v} \times {\bf B} \right]_i \nonumber \\
\thalf \epsilon_{ijk} G_{jk} &=& - \left[ {\bf B} + {\bf v} \times {\bf E}\right]_i
\ea
Defined contravariantly, the relationships between the fields in the two frames of reference are ${\bf E} = \bar{{\bf E}}$, ${\bf B} = \bar{{\bf B}} - {\bf v} \times \bar{{\bf E}}$, and $\bar{{\bf B}} = {\bf B} + {\bf v} \times {\bf E}$.  If instead the fields are defined in terms of the covariant field strength tensor the relationships between them in the two frames of reference are
${\bf B} = \bar{{\bf B}}$, ${\bf E} = \bar{{\bf E}} + {\bf v} \times \bar{{\bf B}}$, and $\bar{{\bf E}} = {\bf E} - {\bf v} \times {\bf B}$.

Expressions for the electric and magnetic fields in terms of the vector potential are more complicated than in an inertial frame, being
\ba
{\bf E} &=& - \left( \frac{\partial}{\partial t} - {\bf v} \cdot \vec{\nabla} \right) \vec{\phi} - 
\left[ \vec{\nabla}  + {\bf v} \left( \frac{\partial}{\partial t} - {\bf v} \cdot \vec{\nabla} \right) \right] \phi^0 \nonumber \\
{\bf B} &=& \vec{\nabla} \times \vec{\phi} + {\bf v} \times \left( \frac{\partial}{\partial t} - {\bf v} \cdot \vec{\nabla} \right) \vec{\phi}
\label{EBphi}
\ea

\section{Field Equations}
\label{secIII}

Consider the classical equations of motion in the inertial frame.  They are
\ba
\bar{G}^{ab} &=& \partial^a \bar{\phi}^b - \partial^b \bar{\phi}^a \nonumber \\
\partial_a \bar{G}^{ab} &=& - m^2 \bar{\phi}^b
\ea
Thus $\partial_b \bar{\phi}^b =0$ which is consistent with three spin degrees of freedom.  Since $\bar{G}^{ab}$ is an antisymmetric tensor, 
$\partial_a \bar{G}^{ab} = {\cal D}_a \bar{G}^{ab}$, where ${\cal D}_a$ is the covariant derivative.  Transformation to the rotating frame leads to
\be
{\cal D}_{\mu} G^{\mu\nu} = \partial_{\mu} G^{\mu\nu} = - m^2 \phi^{\mu}
\label{Gomegaeqs}
\ee
It is also true that
\be
\partial_{\mu} \phi^{\mu} = \partial_a \bar{\phi}^a = 0
\label{phiconstrain}
\ee

Equations (\ref{Gomegaeqs}) can be written in terms of the vector electric and magnetic fields as
\ba
\vec{\nabla} \cdot {\bf E} &=& - m^2 \phi^0 \nonumber \\
\vec{\nabla} \times {\bf B} - \frac{\partial {\bf E}}{\partial t} &=& - m^2 \vec{\phi}
\label{Max12}
\ea
The Bianchi identity 
\be
\partial^{\alpha} G^{\beta\gamma} + \partial^{\gamma} G^{\alpha\beta} + \partial^{\beta} G^{\gamma\alpha} = 0
\ee
is immediately satisfied if one uses $G^{\mu\nu} = \partial^{\mu} \phi^{\nu} - \partial^{\nu} \phi^{\mu}$.  In terms of the electric and magnetic fields
\be
\vec{\nabla} \cdot {\bf B} + {\bf v} \cdot \left( \frac{\partial}{\partial t} - {\bf v} \cdot \vec{\nabla} \right) {\bf B} = 0
\label{Max3}
\ee
and
\be
\left( \frac{\partial}{\partial t} - {\bf v} \cdot \vec{\nabla} \right) {\bf B} + \vec{\nabla} \times {\bf E} +
{\bf v} \times \left( \frac{\partial}{\partial t} - {\bf v} \cdot \vec{\nabla} \right) {\bf E} = 0 
\label{Max4}
\ee 
Some useful relations used include
\ba
\partial^0 = g^{0\sigma} \partial_{\sigma} &=& \partial_0 - v_x \partial_1 - v_y \partial_2 \nonumber \\
\partial^1 = g^{1\sigma} \partial_{\sigma} &=& - \partial_1 - v_x \partial_0 + v_x^2 \partial_1 + v_x v_y \partial_2 \nonumber \\
\partial^2 = g^{2\sigma} \partial_{\sigma} &=& - \partial_2 - v_y \partial_0 + v_x v_y \partial_1 + v_y^2 \partial_2 \nonumber \\
\partial^3 = g^{3\sigma} \partial_{\sigma} &=& - \partial_3
\label{partials}
\ea
or
\ba
\partial^0 &=& \frac{\partial}{\partial t} - {\bf v} \cdot \vec{\nabla} \nonumber \\
\partial^i &=& - \left[ \vec{\nabla} + {\bf v} \left( \frac{\partial}{\partial t} - {\bf v} \cdot \vec{\nabla} \right) \right]_i
\label{contraderiv}
\ea
and
\be
\partial_{\mu} (\partial^{\mu} X) = g^{\mu\nu} \partial_{\mu} \partial_{\nu} X - \omega^2 ( x \partial_1 + y \partial_2 ) X
\label{partial2}
\ee                                                               

Instead of transforming the equations of motion from the inertial to the rotating frame, consider the equations of motion that follow from the Lagrangian.  Although the Lagrangian is unchanged when expressed in terms of the contravariant and covariant field strength tensors and the field, that is not the case when it is expressed in terms of derivatives of the fields.  It is convenient to replace ordinary derivatives $\partial{_\mu}$ with covariant derivatives ${\cal D}_{\mu}$.  The covariant derivative commutes with the metric tensor (covariant, contravariant, or mixed) and thus commutes with the operation of raising or lowering indices.  The covariant curl is equal to the ordinary curl so that 
${\cal D}^{\mu} \phi^{\nu} - {\cal D}^{\nu} \phi^{\mu} = \partial^{\mu} \phi^{\nu} - \partial^{\nu} \phi^{\mu} = G^{\mu\nu}$.  The covariant divergence is equal to the ordinary divergence ${\cal D}_{\mu} \phi^{\mu} = \partial_{\mu} \phi^{\mu}$ because $\det(g_{\mu\nu}) = -1$ is a constant.  Specifically
\be 
{\cal L} = \thalf (\partial_{\alpha} \phi^{\beta}) (\partial_{\beta} \phi^{\alpha}) -
\thalf g_{\alpha\beta} \, g^{\gamma\rho} (\partial_{\gamma} \phi^{\alpha}) (\partial_{\rho} \phi^{\beta}) +
\thalf m^2 g_{\alpha\beta} \, \phi^{\alpha} \phi^{\beta}
\ee
The momentum conjugate to $\phi^{\mu}$ is
\be
\pi^{\mu} = g^{\mu\nu} \frac{\partial {\cal L}}{\partial(\partial_0 \phi^{\nu})}
\ee
As usual one finds that $\pi^0 = 0$ so that $\phi^0$ is not an independent field.  Also $\pi^1 = E_x$ etc. with ${\bf E}$ as given in (\ref{EBphi}).  Thus ${\bf E}$ is the momentum conjugate to $\vec{\phi}$.  

The field equations
\be
g^{\sigma\nu} \left[ {\cal D}_{\mu} \frac{\partial {\cal L}}{\partial(\partial_{\mu} \phi^{\nu})} - \frac{\partial {\cal L}}{\partial \phi^{\nu}} \right] 
= g^{\sigma\nu} \left[ {\cal D}_{\mu} \left( \partial_{\nu} \phi^{\mu} - g_{\alpha\nu} \partial^{\mu} \phi^{\alpha} \right)
- m^2 g_{\alpha\nu} \phi^{\alpha} \right] = 0
\ee
can be put in the form
\be
{\cal D}_{\mu} \left( \partial^{\mu} \phi^{\sigma} - \partial^{\sigma} \phi^{\mu} \right) + m^2 \phi^{\sigma}
= \partial_{\mu} \left( \partial^{\mu} \phi^{\sigma} - \partial^{\sigma} \phi^{\mu} \right) + m^2 \phi^{\sigma} = 0
\ee
consistent with Eq. (\ref{Gomegaeqs}).  Note that the constraint (\ref{phiconstrain}) is automatically satisfied.  Using Eqs. (\ref{partials}) and (\ref{partial2}) the results of the Lagrangian approach are
\ba
g^{\mu\nu} \partial_{\mu} \partial_{\nu} \phi^0 + m^2 \phi^0 &=& \omega^2 (x \partial_1 + y \partial_2) \phi^0
+ \omega (\partial_1 \phi^2 - \partial_2 \phi^1) \nonumber \\
g^{\mu\nu} \partial_{\mu} \partial_{\nu} \phi^1 + m^2 \phi^1 &=& \omega^2 (x \partial_1 + y \partial_2) \phi^1
+ \omega \partial_0 \phi^2 + \omega^2 y (\partial_1 \phi^2 - \partial_2 \phi^1)
 - \omega^2 (x \partial_2 - y \partial_1) \phi^2 \nonumber \\
g^{\mu\nu} \partial_{\mu} \partial_{\nu} \phi^2 + m^2 \phi^2 &=& \omega^2 (x \partial_1 + y \partial_2) \phi^2 
-\omega \partial_0 \phi^1 - \omega^2 x (\partial_1 \phi^2 - \partial_2 \phi^1)
+ \omega^2 (x \partial_2 - y \partial_1) \phi^1 \nonumber \\
g^{\mu\nu} \partial_{\mu} \partial_{\nu} \phi^3 + m^2 \phi^3 &=& \omega^2 (x \partial_1 + y \partial_2) \phi^3
\label{Lphieqs}
\ea
with
\be
g^{\mu\nu} \partial_{\mu} \partial_{\nu} = \partial^2_t - \nabla^2 + v_x^2 \partial_x^2 + v_y^2 \partial_y^2
-2 (v_x \partial_x + v_y \partial_y) \partial_t + 2 v_x v_y \partial_x \partial_y
\ee
Notice the rotational symmetry about the $z$ axis in the equations for the field: $(x,y) \rightarrow (y, -x)$ and $(\phi^1,\phi^2) \rightarrow (\phi^2, -\phi^1)$.  It can be verified that these equations are consistent with the constraint Eq. (\ref{phiconstrain}).

Consider plane wave solutions to Eq. (\ref{Lphieqs}) close to the origin where $|v_x|, |v_y| \ll 1$.  Then it is only necessary to keep the terms of order $\omega$ on the right side of these equations.  Considering the equations for the dynamical components of the field, there is one mode with $E^2 = p^2+m^2 \equiv E_p^2$ and a pair of modes with $E^2 = E_p^2 + \thalf \omega^2 \pm \omega \sqrt{E_p^2 + \oneqt \omega^2}$, for which the positive energies are 
$E = \sqrt{E_p^2 + \oneqt \omega^2} \pm \thalf \omega$.

\section{Determination of the Hamiltonian}
\label{secIV}

In this section we consider a Schr\"odinger-like formulation which involves a Hamiltonian and wave equations with only first order derivatives in time.  For this one needs to make a choice of how to define the wave functions in terms of the fields.  This choice ought to be informed by the requirement that the positive and negative energy states be clearly separated for a particle at rest and, perhaps, with zero vorticity.  There are three natural choices.  The first one is 
\be
\psi_{i\pm} = \thalf \left( \phi^i \pm \frac{i}{m} \frac{\partial \phi^i}{\partial t} \right)
\ee
where $i=1,2,3$ since, for a particle at rest and with zero vorticity, the fields would have the time dependence $e^{-imt}$ for positive energy states and $e^{imt}$ for negative energy states.  However, with vorticity this choice is not the most natural and we do not report on it here.  The second one is
 \be
\psi_{i\pm} = \thalf \left( \phi^i \mp \frac{i}{m} E_i \right) = \thalf \left( \phi^i \mp \frac{i}{m} (\partial^i \phi^0-\partial^0 \phi^i) \right)
\ee
which has the same benefits as the first one with the added bonus that it is a linear combination of the fields and their conjugate momenta~\cite{Silenko2018}.  The resulting Hamiltonian is presented in Appendix B.  The third one is
\be
\psi_{i\pm} = \thalf \left( \phi^i \pm \frac{i}{m} (\partial_t - {\bf v} \cdot \vec{\nabla})\phi^i \right)
\label{psi3}
\ee
which is motivated by the expressions for the electric and magnetic fields (\ref{EBphi}) and the appearance of the contravariant derivatives (\ref{contraderiv}).  Although in the end it should not matter what choice is made, we have found that the third one is the simplest and easiest to work with.

The equations for the independent fields can be written compactly in matrix form as
\be
\begin{split}
      \begin{pmatrix} 
      D - \omega v_x \partial_y && -T + \omega v_x \partial_x && 0\\
      T - \omega v_y \partial_y && D + \omega v_y \partial_x && 0\\
       0 && 0 && D
      \end{pmatrix}     
      \begin{pmatrix}   \phi^1 \\  
                       \phi^2 \\
                       \phi^3
      \end{pmatrix}
=0
\end{split}
\label{matrixphi}
\ee
where
\ba
D &=& g^{\mu\nu} \partial_{\mu} \partial_{\nu} + m^2 - \omega^2 (x \partial_x + y \partial_y) \nonumber \\
 &=& \left( \partial_t - {\bf v} \cdot \vec{\nabla} \right)^2 - \nabla^2 + m^2 \nonumber \\
T &=& \omega (\partial_t - {\bf v} \cdot \vec{\nabla} )
\label{defTD}
\ea
Note that the field $\phi^3$ decouples from the other two.

We focus on the transverse directions first.  Combining Eqs. (\ref{psi3}-\ref{defTD}) we can easily write the exact equations of motion in the form
\be
\begin{split}
i \frac{\partial}{\partial t} \begin{pmatrix}   \psi_{x+} \\  
                       \psi_{y+} \\
                       \psi_{x-} \\
                       \psi_{y-}
      \end{pmatrix} =&       
      H_{\perp} 
      \begin{pmatrix}   \psi_{x+} \\  
                       \psi_{y+} \\
                       \psi_{x-} \\
                       \psi_{y-}
      \end{pmatrix}
\end{split}
\ee
with
\be
H_{\perp} = m \beta +  i {\bf v} \cdot \vec{\nabla} +
\begin{pmatrix} 
  - \dfrac{\nabla^2}{2m} - \dfrac{\omega}{2} \sigma_2 && - \dfrac{\nabla^2}{2m} + \dfrac{\omega}{2} \sigma_2 \\
\\
\dfrac{\nabla^2}{2m} + \dfrac{\omega}{2} \sigma_2 && \dfrac{\nabla^2}{2m} - \dfrac{\omega}{2} \sigma_2\\
\end{pmatrix}     
+ W 
\ee
where
\be
W =
\begin{pmatrix} 
\rm{w} && \rm{w} \\
-\rm{w} && -\rm{w} \\
\end{pmatrix}     
\ee
and
\be
\rm{w}= \dfrac{\omega^2}{2m}
\begin{pmatrix} 
y \partial_y && -y \partial_x  \\
- x \partial_y && x \partial_x
\end{pmatrix}
\label{eq:W}
\ee
Making the identification ${\bf p} = - i \vec{\nabla}$ we see that the entries in $W$ are
\bd
\pm i \frac{\hbar \omega^2}{2 m c^2} x_i p_j
\ed
where $i,j = 1,2$ and factors of Planck's constant and the speed of light have been inserted.  The factor of $i$ is puzzling but, due to the factor of $\hbar$, this term does not enter a classical Hamiltonian.  Due to the factor of $1/c^2$ it vanishes in the nonrelativistic limit.

The Hamiltonian is not Hermitian. However, one can develop a physical quantum theory from a non-Hermitian Hamiltonian if it possesses combined parity ${\cal P}$ and time reversal ${\cal T}$ symmetry. See \cite{Bender1,Bender2} and references therein. Specifically, the energy spectrum of such a Hamiltonian is real and bounded below, the Hilbert space of state vectors is endowed with an inner product having a positive norm, and it generates a unitary time evolution. In order to define a positive and time independent norm for the wavefunction, an additional symmetry must introduced, represented by a linear operator $\mathcal{C}$~\cite{Bender1}, not to be confused with charge conjugation. In addition, to correctly identify the energy spectrum, special care must be taken in specifying the boundary conditions when solving the Schr\"odinger equation. 
Under the combined ${\cal P T}$ symmetry ${\bf x} \rightarrow - {\bf x}$, ${\bf p} \rightarrow {\bf p}$, $ \vec{\omega} \rightarrow - \vec{\omega}$, and $i \rightarrow -i$.  The above Hamiltonian does have this symmetry.  Hence a physical quantum theory can be constructed in a rotating frame.

With a view towards applications to heavy ion collisions, rapidly rotating neutron stars, and rotating atomic gases, we consider an expansion in powers of $\omega$.  One might wish to eliminate the $2 \times 2$ block diagonal terms of the non-Hermitian term $W$, relegating them to higher order in the vorticity, with the transformation $\psi' = e^{iM} \psi$ where in this case $M$ will be time independent.  Then
\be
H' = e^{iM} H e^{-iM} = H + i [M,H] - \thalf [M,[M,H]] + \cdot\cdot\cdot
\ee
We assume 
\be
M =
\begin{pmatrix} 
M_2 + M_1 && 0 \\
0 && M_2 - M_1
\end{pmatrix}
\ee
where
\be
M_1 = \dfrac{\omega}{2m}
\begin{pmatrix} 
A && B \\
C && D
\end{pmatrix}     
\ee
with $A = A_{xx} x \partial_x + A_{xy} x \partial_y + A_{yx} y \partial_x + A_{yy} y \partial_y$, and similarly for $B,\ C,\ D$, a form suggested by Eq.~(\ref{eq:W}), and where
\be
M_2 = - i \dfrac{\omega^2}{2}
\begin{pmatrix} 
a && b \\
c && d
\end{pmatrix}
\ee
with $a = a_1 (x^2 - y^2) + a_2 xy$, and similarly for $b,\ c,\ d$.  The reason for the latter choice is that $\thalf [\nabla^2, a] = 2 a_1 (x \partial_x - y \partial_y) + 
a_2 (y \partial_x + x \partial_y)$.  It is to be understood that all coefficients are dimensionless and independent of $\omega$.  This is a similarity transformation, not a unitary transformation, because we are trying to eliminate a non-Hermitian (but still ${\cal PT}$ symmetric) term in $H$.  Despite having 24 free parameters and 16 equations to solve, no solution can be found as these equations are inconsistent; see Appendix \ref{app:eqsystem}.  We have not discovered any other way to cancel the block diagonal terms in $W$, hence they remain. 

Similarly we may derive the Hamiltonian for the $z$ component
\be
\begin{split}
i \frac{\partial}{\partial t} \begin{pmatrix}   \psi_{z+} \\  
                       \psi_{z-}
      \end{pmatrix} =&       
      H_{||} 
      \begin{pmatrix}   \psi_{z+} \\  
                               \psi_{z-}
      \end{pmatrix}
\end{split}
\ee
to be
\be
H_{||} = m \sigma_3 +  i {\bf v} \cdot \vec{\nabla} +
\begin{pmatrix} 
  - \dfrac{\nabla^2}{2m} && - \dfrac{\nabla^2}{2m} \\
\\
\dfrac{\nabla^2}{2m} && \dfrac{\nabla^2}{2m}
\end{pmatrix}      
\ee
Note the lack of terms that had appeared in the $x$ and $y$ components of the wave function.  This is because the $z$ component represents zero projection of the spin along the vorticity axis.  In principle this Hamiltonian is not Hermitian in the sense that $H^{\dagger} = (H^*)^T \neq H$ on account of the $\nabla^2$ terms.  But this is unrelated to vorticity and always arises with bosons, as has been mentioned many times in the literature.

\section{Foldy--Wouthuysen Nonrelativistic Reduction}
\label{secV}

The leading order spin dependent term calculated at the end of the section III is $\thalf s_z \omega$ where $s_z$ may be identified with the $z$ component of the spin with values 
$0,\pm 1$.  This was derived under the assumptions that the orbital angular momentum is zero and that the vorticity is small, namely, $\omega \ll E_p$.  In this section we perform a Foldy--Wouthuysen nonrelativistic reduction of the field equations.  Such a nonrelativistic reduction for electrically charged vector mesons interacting with the electromagnetic field has been done before; perhaps the first was Ref. \cite{Young}, while a more recent one is Ref. \cite{Silenko2018}.  The tetrads used in our earlier papers \cite{KRR1,KRR3} are still valid if the vorticity is time, but not space, dependent.  Allowing for a time dependence would add additional terms in what we calculate below.  The transformation $\psi' = e^{iS} \psi$, when $S$ is time independent, leads to
\be
H' = e^{iS} H e^{-iS} = H + i [S,H] - \thalf [S,[S,H]] + \cdot\cdot\cdot
\ee

\subsection{Centrifugal and Coriolis forces}

Consider the nonrelativistic reduction for $\psi_{z\pm}$.  The exact Hamiltonian can be written as $H_{||} = m \sigma_3 + {\cal E} + \Omega$ where
\be
{\cal E} = 
      \begin{pmatrix} 
      i {\bf v} \cdot \vec{\nabla} - \dfrac{\nabla^2}{2m} && 0 \\
\\
      0 && i {\bf v} \cdot \vec{\nabla} + \dfrac{\nabla^2}{2m}
      \end{pmatrix}     
\ee
and
\be
\Omega = 
      \begin{pmatrix} 
      0 && - \dfrac{\nabla^2}{2m} \\
\\
      \dfrac{\nabla^2}{2m} && 0
      \end{pmatrix}     
\ee
Now make a unitary transformation with $S = -\frac{i}{2m} \sigma_3 \Omega$ in order to cancel the off-diagonal terms to first order.  Then
\ba
i [S,m \sigma_3] &=& - \Omega \nonumber \\
i [S,{\cal E} + \Omega] &=&  - \frac{(\nabla^2)^2}{4m^3} \left( \sigma_3 + i \sigma_2 \right) 
\nonumber \\
- \thalf [S,[S,m \sigma_3]] &=& \frac{(\nabla^2)^2}{8m^3} \sigma_3
\ea
These make use of the fact that $[{\bf v} \cdot \vec{\nabla}, \nabla^2] = 0$.  This leads to the Hamiltonian for the positive energy states.
\be
H'_{||+} = m - \dfrac{\nabla^2}{2m} + i {\bf v} \cdot \vec{\nabla} - \frac{(\nabla^2)^2}{8m^3} 
\ee
Making the replacement ${\bf p} = - i \vec{\nabla}$ and ${\bf v} = \vec{\omega} \times {\bf r}$ we find
\ba
H'_{||+} &=& m c^2 + \dfrac{{\bf p}^2}{2m} - \vec{\omega} \times {\bf r} \cdot {\bf p} - \frac{({\bf p}^2)^2}{8m^3 c^2} \nonumber \\
&=& m c^2 + \dfrac{{\bf p}^2}{2m} - \vec{\omega} \cdot {\bf L} - \frac{({\bf p}^2)^2}{8m^3 c^2}
\label{relz}
\ea
The third term on the right hand side exactly reproduces the centrifugal and Coriolis forces when using this Hamiltonian to write the classical equations of motion, while the last term is the relativistic correction to the kinetic energy.  

\subsection{Spin effects}
\label{secVspin}

Consider the nonrelativistic reduction for $\psi_{x\pm}$ and $\psi_{y\pm}$.  The $4 \times 4$ Hamiltonian can be written as $H_{\perp} = m \beta + {\cal E} + \Omega$ where
\be
{\cal E} = 
      \begin{pmatrix} 
      i {\bf v} \cdot \vec{\nabla} - \dfrac{\nabla^2}{2m} - \thalf \omega \sigma_2 + \rm{w} && 0 \\
\\
      0 && i {\bf v} \cdot \vec{\nabla} + \dfrac{\nabla^2}{2m} - \thalf \omega \sigma_2 - \rm{w}
      \end{pmatrix}     
\ee
and
\be
\Omega = 
      \begin{pmatrix} 
      0 && - \dfrac{\nabla^2}{2m}  + \thalf \omega \sigma_2 + \rm{w} \\
\\
      \dfrac{\nabla^2}{2m}  + \thalf \omega \sigma_2 - \rm{w} && 0
      \end{pmatrix}     
\ee
As usual we choose $S = -\frac{i}{2m} \beta \Omega$ in order to cancel the off-diagonal block terms in $H_{\perp}$ to order $1/m$.  To this order the term $w$ is not involved.  Then
\ba
i [S,m \beta] &=& - \Omega \nonumber \\
i [S,{\cal E} + \Omega] &=& \dfrac{\nabla^2}{2m^2} \Omega - \frac{1}{m} \left[ \frac{(\nabla^2)^2}{4m^2} - \oneqt \omega^2 \right] \beta
+ \cdot\cdot\cdot \nonumber \\
- \thalf [S,[S,m \beta]] &=& \frac{1}{2m} \left[ \frac{(\nabla^2)^2}{4m^2} - \oneqt \omega^2 \right] \beta + \cdot\cdot\cdot
\ea
This leads to
\be
H'_{\perp} = \left[ m - \dfrac{\nabla^2}{2m} - \frac{(\nabla^2)^2}{8m^3} + \frac{\omega^2}{8m} \right] \beta 
+  i {\bf v} \cdot \vec{\nabla} - \thalf \omega \sigma_2 + \dfrac{\nabla^2}{2m^2} \Omega + W + \cdot\cdot\cdot
\ee
Thus the Hamiltonian to this order for the positive energy states in the given basis is
\be
H'_{\perp +} = m c^2 + \dfrac{{\bf p}^2}{2m} - \vec{\omega} \cdot {\bf L} - \frac{({\bf p}^2)^2}{8m^3 c^2} 
 -\thalf \hbar \omega \sigma_2  + \frac{(\hbar \omega)^2}{8m c^2} + \rm{w}
\label{relxy}
\ee
where $\rm{w}$ is non-Hermitian but ${\cal PT}$ symmetric.

\subsection{Complete spin and relativistic corrections}

Finally we can write the nonrelativistic Hamiltonian for the three independent degrees of freedom, including relativistic corrections, as a $3\times 3$ matrix in the form
\be
H'_{+} = m c^2 + \dfrac{{\bf p}^2}{2m} - \vec{\omega} \cdot {\bf L} - \frac{({\bf p}^2)^2}{8m^3 c^2} 
 -\thalf \hbar \omega S_3  + \frac{(\hbar \omega)^2}{8m c^2} S_3^2 +
\begin{pmatrix} 
      \rm{w} && 0 \\
      0 && 0
      \end{pmatrix} 
\label{relxyz}
\ee
with the spin matrices
\be
S_1 =
\begin{pmatrix} 
      0 && 0 && 0 \\
      0 && 0 && -i \\
      0 && i && 0
      \end{pmatrix} 
= 
      \begin{pmatrix} 
      0 && 0 \\
      0 && \sigma_2
      \end{pmatrix}     
\ee
\be
S_2 =
\begin{pmatrix} 
      0 && 0 && i \\
      0 && 0 && 0 \\
      -i && 0 && 0
      \end{pmatrix}     
\ee
\be
S_3 =
\begin{pmatrix} 
      0 && -i && 0 \\
      i && 0 && 0 \\
      0 && 0 && 0
      \end{pmatrix}
=  
      \begin{pmatrix} 
      \sigma_2 && 0 \\
      0 && 0
      \end{pmatrix}     
\ee
which satisfy $[S_i, S_j] = i \epsilon_{ijk} S_k$ in a standard representation \cite{Young}.

\section{Solution to a Truncated non-Hermitian but ${\cal PT}$ Symmetric Hamiltonian}
\label{secVI}

In this section we solve several truncated versions of the $2 \times 2$ Hamiltonian derived in Sect. \ref{secVspin} to investigate any obvious problems with the non-Hermitian term $\rm{w}$.  First we treat a Hermitian and a non-Hermitian Hamiltonian separately, then we add them together to see if that introduces any complications.  

Consider the Hermitian Hamiltonian $H = - \thalf \omega \sigma_2$ with wavefunction components $\psi_x$ and $\psi_y$.  The energy eigenvalues are $E = \pm \thalf \omega$.  The eigenfunctions are related by $i E \psi_y = -\thalf \omega \psi_x$, but are otherwise unrestricted.  This is elementary quantum mechanics.

Next consider the non-Hermitian but ${\cal PT}$ symmetric Hamiltonian $H = \rm{w}$.  We look for a solution which has rotational symmetry and which is normalizable at $x = y = 0$.  The functional form is 
\ba
\psi_x &=& y f(\rho^2) \nonumber \\
\psi_y &=& - x f(\rho^2)
\ea
where $\rho^2 = x^2 + y^2$.  The pair of coupled scalar equations reduces to
\be
\rho^2 f' = \left( \frac{mE}{\omega^2} - 1 \right) f
\ee
which has solution
\be
f(\rho^2) = \left( \frac{\rho^2}{\rho_0^2} \right)^n
\ee
with
\be
n = \frac{mE}{\omega^2} - 1
\ee
As the integration measure is $d\phi \rho d\rho$, the wavefunction is normalizable at the origin for $E > 0$.  This means the energy spectrum is real and bounded from below. Of course, appropriate boundary conditions must be used at large $\rho$ to avoid the speed of the surface of the rotating cylinder exceeding the speed of light.

Finally we consider the Hamiltonian $H = i {\bf v} \cdot \vec{\nabla} - \thalf \omega \sigma_2 + \rm{w}$ to see if the combination of the non-Hermitian with the Hermitian terms causes any problems.  In this case the solution takes the form
\ba
\psi_x &=& \left(E y + \frac{i \omega}{2} x \right) f(\rho^2) \nonumber \\
\psi_y &=& - \left(E x - \frac{i \omega}{2} y \right) f(\rho^2)
\ea
where $f$ satisfies the differential equation
\be
E^2 f = \omega^2 \left( \frac{E}{m} + \frac{1}{4} \right) f + \frac{\omega^2 E}{m} \rho^2 f'
\ee
The solution is
\be
f(\rho^2) = \left( \frac{\rho^2}{\rho_0^2} \right)^n
\ee
with
\be
n = \frac{mE}{\omega^2} - \frac{m}{4E} - 1
\ee
The solution is normalizable at the origin if either $-\thalf \omega < E < 0$ or if $E > \thalf \omega$.  It is interesting that there is a gap in the spectrum.  Nevertheless, it seems that the non-Hermitian but ${\cal PT}$ symmetric term $\rm{w}$ results in real energy eigenvalues bounded from below.

\section{Conclusion}
\label{secVII}
 In this paper, motivated by the observation of vorticity in the quark-gluon plasma produced in non-central heavy ion collisions at RHIC and LHC, we investigated the coupling of spin and vorticity of massive vector mesons in a rotating frame of reference.  Starting from the Proca equations of motion in non-inertial frames, we derived the Hamiltonian in a Schr\"odinger-like formulation. We found this Hamiltonian to be non-Hermitian but $\mathcal{PT}$ invariant. We found the vorticity dependent non-Hermitian term in the Hamiltonian to be both a relativistic and quantum correction $\mathcal{O}(\hbar/c^2)$. We recover the nonrelativistic Coriolis and centrifugal forces from the Foldy-Wouthuysen transformation, and obtained the Hamiltonian for the positive energy states, including leading relativistic corrections. There is a splitting of $\frac{1}{2}s_z\omega$ to leading order in the vorticity.

\section*{Acknowledgement}
The work of JIK was supported by the U.S. Department of Energy Grant DE-FG02-87ER40328.  The work of ER was supported by the U.S. National Science Foundation Grant PHY-1630782 and by the Heising-Simons Foundation Grant 2017-228.

\appendix

\section{Metric}

Consider a region of space where a fluid element is rotating in an anti-clockwise sense around the $z$ axis with angular speed $\omega$ which may be considered constant within that region.  We choose the tetrad as the $4\times 4$ matrix
\be
e_{\mu}^{\;\;a}(x) =
\begin{pmatrix}
1 & v_x & v_y & 0 \\
0 & 1 & 0 & 0 \\
0 & 0 & 1 & 0 \\
0 & 0 & 0 & 1 \\
\end{pmatrix}
\ee
where $v_x \equiv -\omega y$ and $v_y \equiv \omega x$.  From this is it straightforward to find the metric
\be
g_{\mu\nu}(x) =
\begin{pmatrix}
1 -v^2 & -v_x & -v_y & 0 \\
-v_x & -1 & 0 & 0 \\
-v_y & 0 & -1 & 0 \\
0 & 0 & 0 & -1 \\
\end{pmatrix} \, ,
\ee
the inverse metric
\be
g^{\mu\nu}(x) =
\begin{pmatrix}
1 & -v_x & -v_y & 0 \\
-v_x & -1+v_x^2 & v_x v_y & 0 \\
-v_y & v_x v_y & -1+v_y^2 & 0 \\
0 & 0 & 0 & -1 \\
\end{pmatrix} \, ,
\ee
and the inverse tetrad
\be
e^{\mu}_{\;\;a}(x) =
\begin{pmatrix}
1 & 0 & 0 & 0 \\
-v_x & 1 & 0 & 0 \\
-v_y & 0 & 1 & 0 \\
0 & 0 & 0 & 1 \\
\end{pmatrix} \, .
\ee
The nonzero components of the affine connection are
\ba
\Gamma^1_{00} &=& \omega v_y \nonumber \\
\Gamma^2_{00} &=& -\omega v_x \nonumber \\
\Gamma^2_{01} &=& \omega \nonumber \\
\Gamma^1_{02} &=& -\omega \,.
\ea

\section{Hamiltonian for an alternate choice of wavefunction}

As the Hamiltonian derived in this work is non-Hermitian one is left to ponder whether alternative wavefunctions can be found that lead to the usual quantum theory we are familiar with.  In this appendix we consider the choice made in Ref. \cite{Silenko2018}, namely
\be
\psi_{i \pm} = \thalf \left( \phi^i \mp \frac{i}{m} E_i \right) = \thalf \left( \phi^i \mp \frac{i}{m} (\partial^i \phi^0-\partial^0 \phi^i) \right)
\ee
which is a linear combination of the fields and their conjugate momenta.  The resulting Hamiltonian is the $6 \times 6$ matrix operator
\be
H =
\begin{pmatrix} 
      m + h_0 + h_2  && h_1 + h_3\\
      h_1 - h_3 && -m + h_0 - h_2 
      \end{pmatrix}     
\ee
where
\be
\begin{split}
h_0 =& \frac{i}{2}\bigg[ 2({\bf{v}} \cdot \vec{\nabla}) - [{\bf S}\cdot  ({\bf{S}} \cdot {\bf{v}})\vec{\nabla}]+ ({\bf{S}} \cdot \vec{\nabla})({\bf{S}}\cdot {\bf{v}})\\
&-\frac{1}{ m^2}({\bf{S}} \cdot \vec{\nabla})({\bf{S}}\cdot {\bf{v}})\left[1+({\bf{S}}\cdot {\bf v})^2 - {\bf{v}}^2\right]  
\left[\nabla^2 -({\bf{S}}\cdot \vec{\nabla})^2\right]\bigg]\\
h_1 =& \frac{i}{2}\bigg[ 2({\bf{v}} \cdot \vec{\nabla}) - [{\bf S}\cdot  ({\bf{S}} \cdot {\bf{v}})\vec{\nabla}]- ({\bf{S}} \cdot \vec{\nabla})({\bf{S}}\cdot {\bf{v}})\\
&+\frac{1}{ m^2}({\bf{S}} \cdot \vec{\nabla})({\bf{S}}\cdot {\bf{v}})\left[1+({\bf{S}}\cdot {\bf v})^2 - {\bf{v}}^2\right]  
\left[\nabla^2 -({\bf{S}}\cdot \vec{\nabla})^2\right]\bigg]\\
h_2=&\frac{1}{2 m} \bigg[({\bf{S}} \cdot \vec{\nabla})({\bf{S}}\cdot {\bf{v}})[{\bf S}\cdot  ({\bf{S}} \cdot {\bf{v}})\vec{\nabla}]-({\bf{S}} \cdot \vec{\nabla})^2 \\
&- \left[1+({\bf{S}}\cdot {\bf v})^2 - {\bf{v}}^2\right]  
\left[\nabla^2 -({\bf{S}}\cdot \vec{\nabla})^2\right]-  ({\bf{S}} \cdot \vec{\nabla})({\bf{S}}\cdot {\bf{v}}) ({\bf{v}} \cdot \vec{\nabla}) \bigg]\\
h_3=& \frac{1}{2m} \bigg[({\bf{S}} \cdot \vec{\nabla})({\bf{S}}\cdot {\bf{v}})[{\bf S}\cdot  ({\bf{S}} \cdot {\bf{v}})\vec{\nabla}]-({\bf{S}} \cdot \vec{\nabla})^2\\
&+\left[1+({\bf{S}}\cdot {\bf v})^2 - {\bf{v}}^2\right]  
\left[\nabla^2 -({\bf{S}}\cdot \vec{\nabla})^2\right]-  ({\bf{S}} \cdot \vec{\nabla})({\bf{S}}\cdot {\bf{v}}) ({\bf{v}} \cdot \vec{\nabla}) \bigg]
\end{split}
\ee
The ${\bf S}$ are the $3 \times 3$ spin matrices as given in the text.  The block off-diagonal terms $h_1 \pm h_3$ couple the positive and negative energy states.  
The term $h_3$ makes this Hamiltonian non-Hermitian. This Hamiltonian does possess $\mathcal{PT}$ symmetry and therefore is acceptable.  
However, we found the Hamiltonian for this choice of wavefunction more complicated because it is third order in derivatives and it couples the $x$ and $y$ components of the wavefunction to the $z$ component.  We do not pursue it in this paper. 

\section{An attempt to remove the non-Hermitian term}
\label{app:eqsystem}

In this appendix we present some details of the results mentioned in Sect. \ref{secIV} for the attempt at removing the non-Hermitian part of the Hamiltonian with a similarity transformation.  The terms needed are
\ba
i [M_1, i {\bf v} \cdot \vec{\nabla}]_{11} &=&  \dfrac{\omega^2}{2m} 
\left\{ (A_{xy} + A_{yx}) (x \partial_x - y \partial_y)  + (A_{yy} - A_{xx}) ( x \partial_y + y \partial_x)  \right\} \nonumber \\
i [M_1, i {\bf v} \cdot \vec{\nabla}]_{12} &=&  \dfrac{\omega^2}{2m} 
\left\{ (B_{xy} + B_{yx}) (x \partial_x - y \partial_y)  + (B_{yy} - B_{xx}) ( x \partial_y + y \partial_x)  \right\} \nonumber \\
i [M_1, i {\bf v} \cdot \vec{\nabla}]_{21} &=&  \dfrac{\omega^2}{2m} 
\left\{ (C_{xy} + C_{yx}) (x \partial_x - y \partial_y)  + (C_{yy} - C_{xx}) ( x \partial_y + y \partial_x)  \right\} \nonumber \\
i [M_1, i {\bf v} \cdot \vec{\nabla}]_{22} &=&  \dfrac{\omega^2}{2m} 
\left\{ (D_{xy} + D_{yx}) (x \partial_x - y \partial_y)  + (D_{yy} - D_{xx}) ( x \partial_y + y \partial_x)  \right\} 
\ea
\ba
i \left[M_1, -\thalf \omega \sigma_2 \right]_{11} &=& \dfrac{\omega^2}{2m} \left\{ \thalf (B+C) \right\} \nonumber \\
i \left[M_1, -\thalf \omega \sigma_2 \right]_{12} &=& \dfrac{\omega^2}{2m} \left\{ \thalf (D-A) \right\} \nonumber \\
i \left[M_1, -\thalf \omega \sigma_2 \right]_{21} &=& \dfrac{\omega^2}{2m} \left\{ \thalf (D-A) \right\} \nonumber \\
i \left[M_1, -\thalf \omega \sigma_2 \right]_{22} &=& \dfrac{\omega^2}{2m} \left\{ - \thalf (B+C)  \right\}
\ea
\ba
i \left[M_2, - \dfrac{\nabla^2}{2m} \right]_{11} &=& 
\dfrac{\omega^2}{2m} \left\{ 2 a_1 (x \partial_x - y \partial_y) + a_2 (y \partial_x + x \partial_y) \right\} \nonumber \\
i \left[M_2, - \dfrac{\nabla^2}{2m} \right]_{12} &=&
 \dfrac{\omega^2}{2m} \left\{ 2 b_1 (x \partial_x - y \partial_y) + b_2 (y \partial_x + x \partial_y) \right\} \nonumber \\
i \left[M_2, - \dfrac{\nabla^2}{2m} \right]_{21} &=&
 \dfrac{\omega^2}{2m} \left\{ 2 c_1 (x \partial_x - y \partial_y) + c_2 (y \partial_x + x \partial_y) \right\} \nonumber \\
i \left[M_2, - \dfrac{\nabla^2}{2m} \right]_{22} &=&
 \dfrac{\omega^2}{2m} \left\{ 2 d_1 (x \partial_x - y \partial_y) + d_2 (y \partial_x + x \partial_y) \right\}
\ea
To cancel the non-Hermitian, order $\omega^2$, term in the original Hamiltonian we need the following equations to hold.\\

\noindent From the 12 component
\ba
\left( B_{xy} + B_{yx} + \thalf D_{xx} - \thalf A_{xx} + 2b_1 \right) x \partial_x &+& \nonumber  \\
\left( B_{yy} - B_{xx} + \thalf D_{xy} - \thalf A_{xy} + b_2 \right) x \partial_y \nonumber &+& \\
\left( B_{yy} -B_{xx} + \thalf D_{yx} - \thalf A_{yx} + b_2 - 1 \right) y \partial_x \nonumber &+& \\
\left( -B_{xy} - B_{yx} + \thalf D_{yy} - \thalf A_{yy} - 2b_1 \right) y \partial_y &=& 0
\label{M12}\ea
From the 21 component
\ba
\left( C_{xy} + C_{yx} + \thalf D_{xx} - \thalf A_{xx} + 2c_1 \right) x \partial_x &+& \nonumber  \\
\left( C_{yy} - C_{xx} + \thalf D_{xy} - \thalf A_{xy} + c_2 - 1 \right) x \partial_y \nonumber &+& \\
\left( C_{yy} -C_{xx} + \thalf D_{yx} - \thalf A_{yx} + c_2 \right) y \partial_x \nonumber &+& \\
\left( -C_{xy} - C_{yx} + \thalf D_{yy} - \thalf A_{yy} - 2c_1 \right) y \partial_y &=& 0
\label{M21}
\ea
The same derivations can be performed for the other components, resulting in 16 scalar equations and 24 parameters. 
Let us focus on the $x \partial_y$ and $y \partial_x$ terms in Eqs. (\ref{M12}) and (\ref{M21}), which are

\ba
B_{yy} - B_{xx} + \thalf D_{xy} - \thalf A_{xy} + b_2  &=& 0 \nonumber \\
B_{yy} -B_{xx} + \thalf D_{yx} - \thalf A_{yx} + b_2 - 1 &=& 0 \nonumber \\
C_{yy} - C_{xx} + \thalf D_{xy} - \thalf A_{xy} + c_2 - 1 &=& 0 \nonumber \\
C_{yy} -C_{xx} + \thalf D_{yx} - \thalf A_{yx} + c_2 &=& 0
\label{M1221}
\ea
The fourth equation leads to
\be
A_{yx} = 2 c_2 -2 C_{xx} + 2 C_{yy} + D_{yx} 
\label{M1221_4}
\ee
while a combination of the first three equations lead to
\be
A_{yx} = 2 c_2-2 C_{xx}+2 C_{yy}+D_{yx}-4
\label{M1221solved}
\ee
which are clearly inconsistent.

\end{document}